\documentclass{article}

\usepackage{PRIMEarxiv}

\usepackage[utf8]{inputenc} 
\usepackage[T1]{fontenc}    
\usepackage{hyperref}       
\usepackage{url}            
\usepackage{booktabs}       
\usepackage{amsfonts}       
\usepackage{nicefrac}       
\usepackage{microtype}      
\usepackage{lipsum}
\usepackage{fancyhdr}       
\usepackage{graphicx}       
\graphicspath{{media/}}     
\usepackage{float}

\title{Vision Controlled Sensorized Prosthetic Hand
}

\author{
   Md Abdul Baset Sarker, Juan Pablo S. Sola,  Aaron Jones, Evan Laing, Ernesto Sola-Thomas, Masudul H. Imtiaz \\
  Clarkson University\\
  Potsdam, NY, USA\\
  \texttt{sarkerm@clarkson.edu, schumaj@clarkson.edu, aarjones@clarkson.edu } \\
  \texttt{lainge@clarkson.edu, schumae@clarkson.edu, mimtiaz@clarkson.edu} \\
}

\begin{document}
\maketitle

\begin{abstract}
This paper presents a sensorized vision-enabled
prosthetic hand aimed at replicating a natural hand’s
performance, functionality, appearance, and comfort. The design
goal was to create an accessible substitution with a user-friendly
interface requiring little to no training. Our mechanical hand uses
a camera and embedded processors to perform most of these tasks.
The interfaced pressure sensor is used to get pressure feedback
and ensure a safe grasp of the object; an accelerometer is used to
detect gestures and release the object. Unlike current EMG-based
designs, the prototyped hand does not require personalized
training. The details of the design, tread-off, results, and
informing the next iteration are presented in this paper.

Video: \href{https://www.youtube.com/watch?v=yInMzl4Ef7A}{https://www.youtube.com/watch?v=yInMzl4Ef7A} 
\end{abstract}


\section{Introduction}
By the year 2050, the number of people estimated to be
living with the loss of an upper limb is 3.6 million \cite{1}. Since the
use of the upper extremities is essential for communication and
the performance of day-to-day activities, assistive devices have
been developed, modified, and researched to aid impaired
individuals. These prosthetics would have the capability to
enable people with functional difficulties to lead more
productive and dignified lives, participating in education, the
labor market, and social life \cite{2}. Prosthetic devices can be
classified into two categories: fully actuated and under-actuated.
A fully actuated prosthetic more closely replicates the actions
and mobility of a natural hand, including up to 27 degrees of
freedom: four in each finger, three for extension and flexion, and
one for abduction and adduction \cite{2} \cite{3}. An under-actuated
prosthetic removes some degrees of freedom to reduce
complexity, cost, power consumption, and weight. Most
prosthetic hands tend towards under-actuated mechanisms,
often by fixing the thumb in place, making it far more difficult
for users to grasp a wide range of objects. 

Current research has emphasized the implementation of
various sensors into the prosthetic hand. Joseph et al. \cite{4}
developed a prosthetic hand by placing a camera in the palm to
perform real-time grasp selection and implementing a
convolutional neural network (CNN) on the NVIDIA Tegra
processor. They took images for training from DeepGrasping,
ImageNet, and locally captured images. Their design utilized a
single-board computer too large to be placed on the hand. 
Another prosthetic hand using cameras was developed by
Shi et al. \cite{5} that can perform “reach-and-pick up” tasks on
various daily objects with vision-based myoelectric control. The
authors used a 3D camera, Kinect2.0, to capture color
(1920×1080) and depth (512 × 424) images at 30 Hz. They
placed the camera at a distance from the table, not on the hand
itself. A multilayer CNN was used to determine the grasp
selection, and an EMG signal was used to control the device. 
Weiner et al. \cite{6} placed a 1.3-megapixel RGB camera on the
palm and an LCD on the black side of their KIT underactuated
prosthetic hand. Using two motors, it could provide 10 degrees
of freedom. No additional sensors were added to provide an
accurate distance measurement; rather, the camera was used to
serve this purpose. An ARM Cortex M7 processor (STM32F7)
was the main processor.

We also based our design on cameras and deep learning but
used more sensors. We added a 5-megapixel camera on the wrist
position for vision and five pressure sensors (FSR) on the
fingertips to get pressure feedback. A distance sensor was also
placed beside the camera to get the accurate distance of the
object. Finally, an accelerometer was added inside the body of
the hand, enabling gesture recognition for object release. Our
motivation was to introduce a 3D printed mechanical hand that
can replicate most of the motions and functions a natural hand
can perform, such as grasping various objects, tactile feeling,
object recognition, and manipulating objects of various shapes. 
Multiple sensors were implemented on our prosthetic hand to
obtain the functionalities.

The paper is organized as follows. First, the design
considerations are presented in Section \ref{sec:sec_design_consideration}. The prototype
description is found in Section \ref{sec:protype_description}. The 3D hand-design
procedure is explained in Section \ref{sec:mechanical_design}. Section \ref{sec:electronic_hardware} presents a
detailed description of the electronic hardware, software, and
sensors. The system validation procedure is provided in Section
\ref{sec:system_validation}. Section \ref{sec:discussion} discusses the validation results, possible
applications, system limitations, and future works. The paper is
concluded in Section \ref{sec:conclusion}.

\section{Design Consideration}
\label{sec:sec_design_consideration}
Our research aim was to develop a prosthetic hand that
would be an accessible substitution requiring little to no training
and providing the maximum Degrees of Freedom. It was
necessary to have good processing power to run vision-enabled
models on an embedded processor while maintaining low power
consumption. We chose the Coral Dev Board Mini because it is
a compact (64mm X 48mm) low-powered single-board
computer. Another consideration was based on study results
showing that amputees want to reduce the attention they need to
invest to use a prosthetic to grasp an object \cite{6,7,8} and the time
required for training. EMG signal is widely used in prosthetic
devices \cite{9,10,11}. The quality of this EMG interface varies due
to placement and user conditions \cite{6} and may be ineffective for
those with upper-arm disabilities. Our design focuses on using
the vision-based solution rather than EMG for primary control.
For object detection, our device uses CNN, which is popular
among researchers \cite{12,13,14,15,16,17,18,19,20,21}. To perform object detection in the
embedded device, we have chosen EfficientDet, a scalable and
efficient object detection model \cite{22} that can run on embedded
devices like Coral Dev Board Mini \cite{23}.

Studies have shown that using an elastic or viscoelastic
material on the palm and fingers is beneficial for grasping
objects due to their increased surface area upon contact \cite{3}.
However, products using these materials tend to be more
expensive, making them less accessible to many users. Indeed,
prosthetics have always been considered costly due to their
maintenance cost. We chose PLA (Polylactic acid) to print the
hand to make the device lightweight, strong, and affordable.
PLA is lightweight, low-cost, and used in different medical and
pharmacological applications for 3D object printing \cite{24}.

\section{Prototype Description}
\label{sec:protype_description}
\begin{figure}[H]
  \centering
  \includegraphics[width=0.6\textwidth]{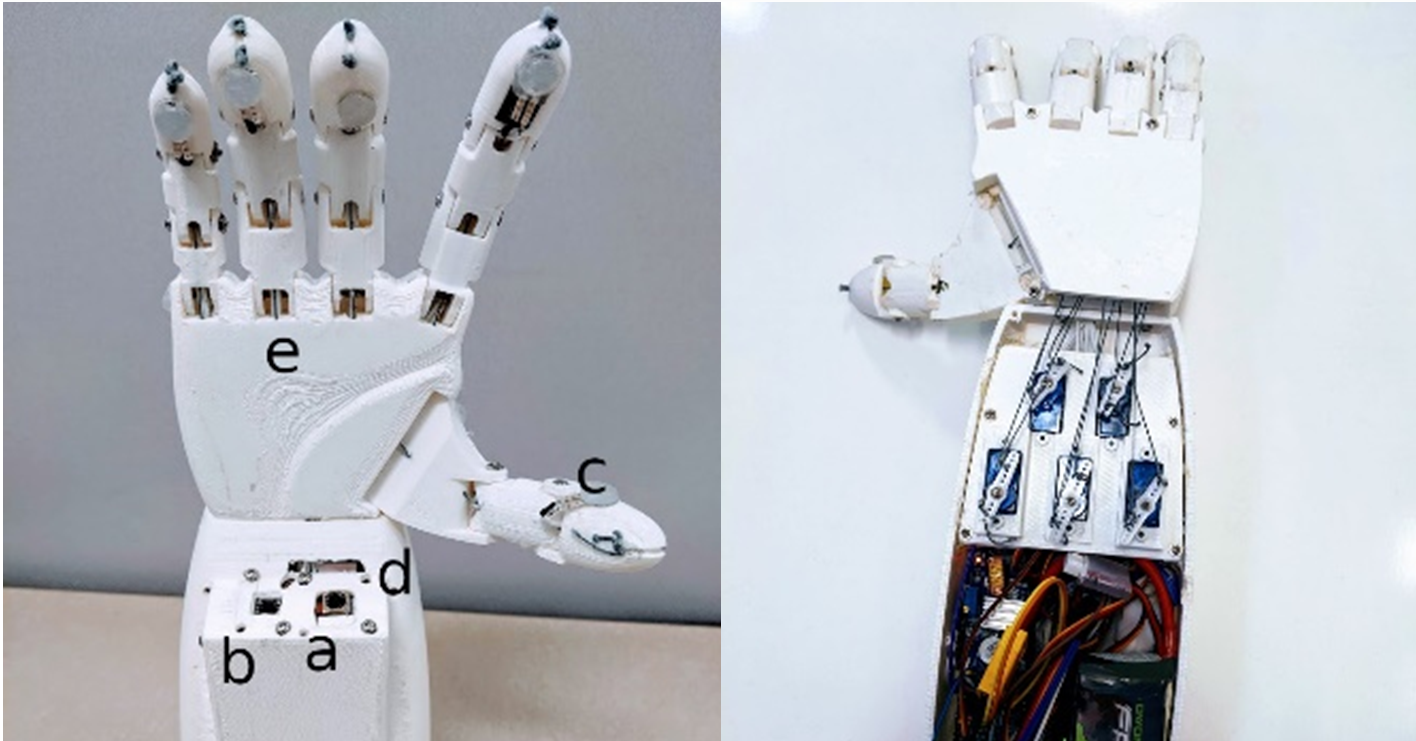}%
  \caption{(left) shows the front side of the working prototype of Vision Controlled Sensorized Prosthetic hand: (a) Camera, (b) Distance sensor, (c) Pressure sensor (FSR), (d) accelerometer (hidden inside),(e) Tendon wire. and the image (right) is the backside showing the electronic circuit and servo motor assembly}
  \label{fig:fig1}
\end{figure}

Figure \ref{fig:fig1} shows the working prototype of the hand and its
major sensors.The main processor and other electronic
components are housed inside the forearm. This design includes
five FSRs, one camera, and one accelerometer. FSRs are placed
on the fingertips to provide tactical feedback
Each FSR will provide feedback with which the system can determine the force
at which an object is being held. The camera is located on the
wrist to implement and facilitate automatic object recognition
with the help of the proximity distance sensor. The design and
firmware are made open-source (hosted in GitHub \cite{25} to
facilitate mass community adoption. All the sensors described
above are interfaced with a single-board computer which
collects sensor data from the peripherals and provides control
signals to the hand motors to perform the required movements
for the prosthetic. 

The total length of the designed prosthetic hand after
assembly is 373mm, and the size of the palm is “80mm x 86mm
x 33.5mm(LxWxD)”. The middle finger length is 88.3mm. The
size of the forearm is “205.00mm x 90.50mm x 80mm
(LxWxD)”.

\section{Mechanical Design}
\label{sec:mechanical_design}
\begin{figure}[H]
  \centering
  \includegraphics[width=0.60\textwidth]{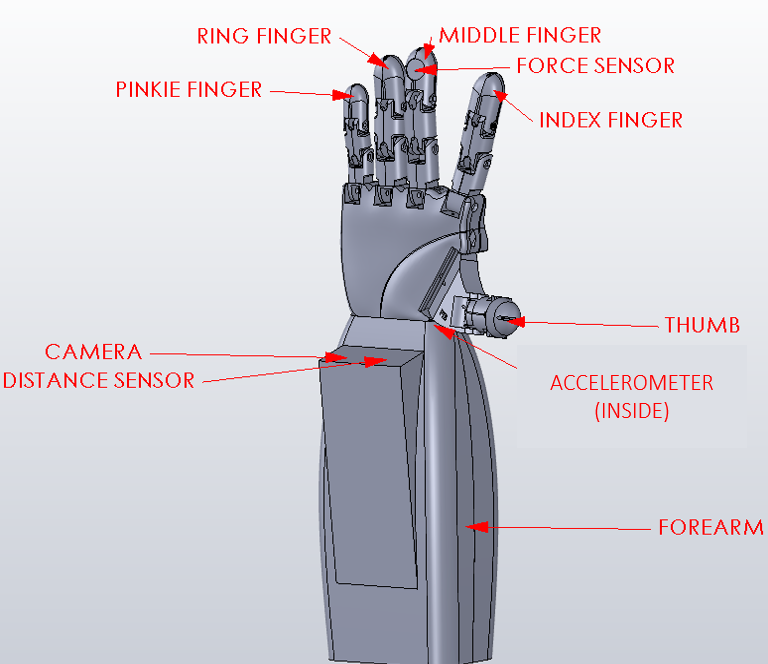}%
  \caption{3D Design of Vision-Enabled Sensorized Prosthetic Hand}
  \label{fig:fig2}
\end{figure}

While designing the hand, we considered the system’s
strength, reliability, and mobility, as well as the placement and
space available for a wide array of sensors and control devices
for the hand. The prototype shown below (Figure \ref{fig:fig2}) was printed
using Ender 5 3D printer, consuming approximately 289 grams
of PLA plastic over around 30 hours.
Each finger has its own servo motor. Two tendon wires
connect each finger to its servo motor, providing two-directional 
movement. Our prototype design used the SG90 9G servo,
which is robust and readily available.

\section{Electronic Hardware}
\label{sec:electronic_hardware}
\begin{figure}[H]
  \centering
  \includegraphics[width=0.90\textwidth]{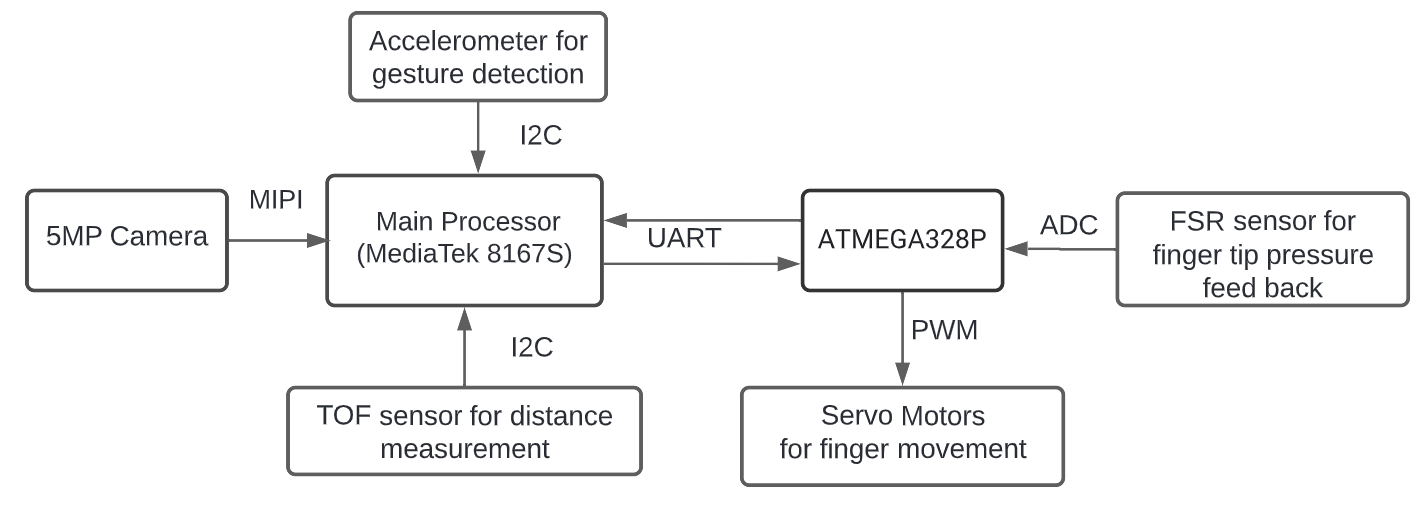}%
  \caption{Block diagram of the system}
  \label{fig:fig3}
\end{figure}

Figure \ref{fig:fig3} shows the block diagram of the system, where a
MediaTek 8167S (Coral Dev Board Mini) is used as the main
processor. A 5MP camera, accelerometer, Timeof-Flight sensor, and a Motor Controller are directly connected
to this main processor. A motor controller ATMEL328P is also
connected to the five pressure sensors and servo motors. The
pressure sensors are placed on the fingertips of the 3D-printed
hand. The camera is connected through the MIPI interface. The
accelerometer and TOF sensor are connected to the main
processor via an I2C connection. The main processor and
ATMEL328P are connected through UART. Pressure sensors
(FSR) are connected to ATMEL328P using a 10-bit analog-to-digital converter (ADC). Finally, the servo motors are
controlled over a PWM signal from five digital I/O pins of
ATMEGA328P. We used 11.1v lithium-ion battery with a buck
converter that converts 5v to run the devices. 

\subsection{Coral dev board mini}
Coral Dev Board Mini is a low-cost single-board computer.
The small version provides fast machine learning (ML)
inference \cite{23}. The USB-C-powered Coral Dev Board Mini
has a CPU (MediaTek 8167S), GPU (IMG PowerVR GE8300),
2 GB LPDDR3 RAM, 8 GB eMMC flash memory, and more. It
has wireless connectivity like Bluetooth and Wi-Fi on board
\cite{23}. It also includes an Edge Tensor Processor Unit (TPU) \cite{26}.
The Edge TPU has outperformed similar form factor devices’
for both latency and computational efficiency \cite{27,28}. The
performance increases dramatically when Edge TPU is included
compared to non-Edge TPU devices \cite{28}. Furthermore, it
provides good performance with low power consumption \cite{29}.
With Coral Dev Board Mini, TensorFlow Lite models can be
compiled and run on the Edge TPU. We installed the Mendel Operating system in the internal
8GB flash memory of the Coral Dev Board Mini. A Python
script was developed to run continuously upon startup and
complete the following tasks: 

\begin{itemize}
\item Upon object detection, an accurate distance is
calculated using the TOF sensor (VL6180x).
\item If the object is near to the hand, about (70-90mm),
it sends a signal “Hand Close” to the motor
controller to close the hand.
\item While the hand is closing, it gets feedback from the
pressure sensors. If the pressure value exceeds a
predefined value, it will stop closing the hand.
\item Once the hand has closed, the accelerometer
(ADXL345) is activated and waits to detect a
certain gesture. Upon sensing the gesture, it will
send the “Hand Open” signal to the motor
controller.
\end{itemize}

\subsection{Camera and Sensors }

A 5-megapixel camera module (25mm x 25mm) compatible
with Coral boards is used to capture the images. The camera
connects through the MIPI-CSI interface and provides an easy
way to bring visual input into the model. The camera has
autofocus with a focal length of 2.5mm and a range of 10cm to infinity. 
The camera’s view angle is 84/87 degrees, and it has
auto white balance control, auto exposure control, auto band
filter, auto black level calibration, and auto 50/60Hz
illumination that helps get optimized images for object detection
\cite{30}. Some researchers placed the camera inside the thumb \cite{6}.
we decided to place the camera on the wrist to reduce interference from the
fingers.
Five Force Sensitive Resistors (FSRs) are attached to each
fingertip of the prosthetic to get pressure feedback for the hand
while grasping the object. When the “Hand Close” command is
initiated, the motor controller starts to close the hand by
controlling servo motors by sensing the pressure from FSR.
The VL6180X is a Time of Flight distance sensor that
contains a very tiny laser source and a matching sensor. The
VL6180X can detect the Time of Flight for the laser to bounce
back to the sensor; VL6180X uses a very narrow light source to
determine the distance of the surface directly in front of it. The
VL6180X is much more precise and doesn’t have linearity or
‘double imaging’ problems compared to other solutions \cite{31}.
VL6180X is used here to detect the exact distance of the object
from the hand. When an object is detected by the main camera,
VL6180X is activated and calculates the object’s distance. The
main processor sends a signal to close the hand if the object is
between 70-90mm away.
The ADXL345 is a low-power, 3-axis MEMS (MicroElectro-Mechanical Systems) accelerometer module. The
MEMS dimension can range from several millimeters to less
than one micrometer. The MEMS consists of a micro-machined
structure on a silicon wafer suspended by polysilicon springs
that allow it to deflect smoothly on the X, Y, or Z-axis \cite{32}.
Between the plates, the defection causes a measurable change in
capacitance. \cite{33}. ADXL345 is a triple-axis accelerometer with
a Digital I2C and SPI interface breakout. There is an onboard
3.3V regulator and logic-level shifting, so it is easy to interface
with a 3V or 5V microcontroller. The sensitivity level can be set
to either +-2g, +-4g, +-8g, or +-16g, where the lower range gives
more resolution for slow movements, and the higher range is
good for high-speed tracking. ADXL345 is used in our
prosthetic hand to detect a gesture (tilt) to release the object. 

\subsection{ Motor Controller and Motors}
ATMEGA328P microcontroller is used as the motor
controller. Atmega328p is an 8-bit RSC-based microcontroller
that combines 32KB flash memory, 1024B EEPROM, 2KB 
RAM, 32 I/O pins, 6-channel 10-bit A/D converter, UART, SPI,
etc. \cite{34}. The ATMEGA328P is connected to the main
processor through serial communication that accepts the signals
“Hand Close” and “Hand Open.” Five SG90 micro servo motors
manipulate the hand via tendon wires. 

\subsection{Training and object detection}
As the Coral Dev Board Mini has a TPU (Tensor
Processing Unit) \cite{35}, we have used the EfficientDet-Lite0
model architecture \cite{23}, \cite{36}. For training, we collected
numerous images of two objects. We annotated all the images
using the open-source LabelImg script \cite{37}. Then augmented
the images and separated 10\% used for testing, 10\% for
validation, and 80\% for training. 

\section{System Validation}
\label{sec:system_validation}
As it is a vision-controlled device, our first target was to get
real-time object detection. The device was trained and tested
with the custom objects. For testing, two objects of different
colors were trained and tested with the grasp, lift, and drop
functionality several times. These actions were repeated in
various locations and lighting conditions. A video was published
on YouTube to demonstrate the working of the Vision-Enabled
Prosthetic Hand \cite{38}. We ran the test 20 times and got 90\%
accuracy.
As this is a battery-operated device, one of the key
requirements is to ensure that a universal and commercially
available battery charging system can provide the required
current \cite{39}. In this project, we have used one 1300mAh
Lithium-Polymer battery. A commercially available lithium polymer charger is used to charge the battery. Any 11.1v charger
can charge these batteries. Under lab conditions, this device ran up to one hour under constant use with a single charge. 

\begin{figure}[H]
  \centering
  \includegraphics[width=0.60\textwidth]{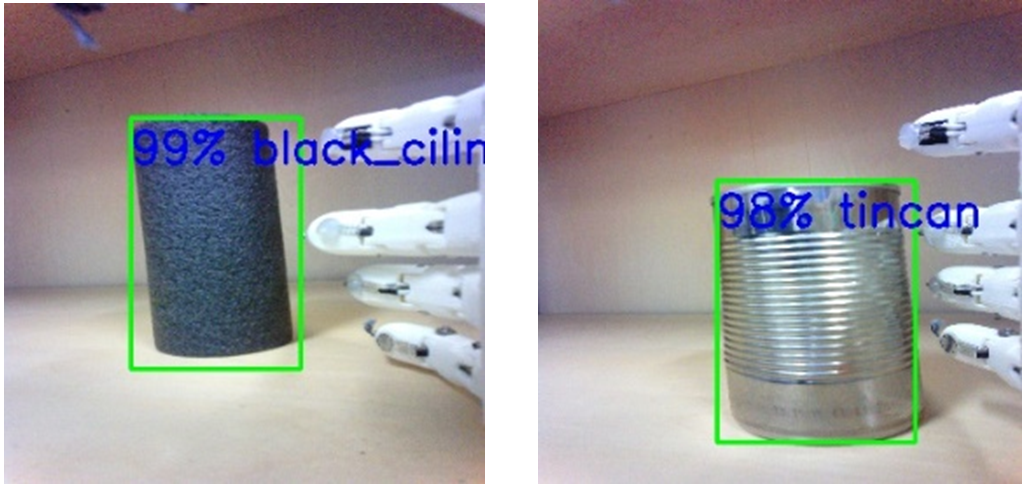}%
  \caption{Object detected when the hand is closer to the object}
  \label{fig:fig4}
\end{figure}

\begin{figure}[H]
  \centering
  \includegraphics[width=0.90\textwidth]{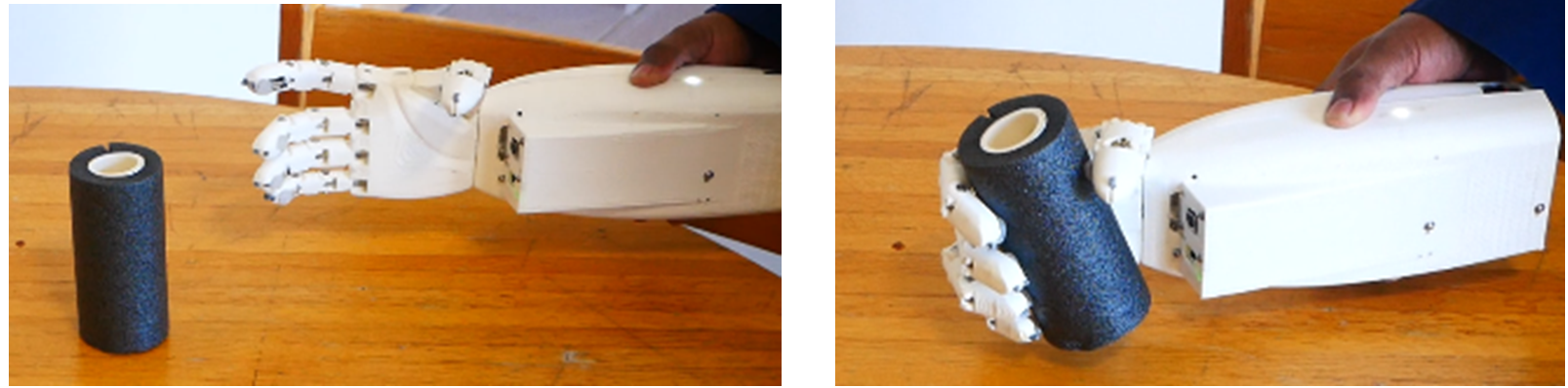}%
  \caption{Grasping the object with the prosthetic hand}
  \label{fig:fig5}
\end{figure}

\section{Discussion}
\label{sec:discussion}
In this study, we designed a prosthetic hand with increased
functionality, lower weight, and adequate application of force
to grasp an object. One of our primary design goals was to
reduce the weight and cost, which led to the decision to 3D print
the hand using low-cost, durable, and lightweight PLA
material. The total weight of the prosthetic hand, including the
motor, battery, electronics, and structure, was 413g. For object
detection, EfficientDet was used in a Coral Dev Board Mini.
The image was captured using a 5MP Coral camera and resized
to 320X320 pixels. The pre-trained weights were loaded into
the device’s flash memory and performed continuously at 9
FPS. Figure \ref{fig:fig4} shows the object detection when the object is
close to the hand. It shows that the tin can and the black cylinder
is detected by the hand, and it’s ready to grasp the object.

The battery voltage is 11.1v. While idle, the whole system
draws 130 mA current. Running the inference takes around
250mA current, and while in operation (object detection and
grasping an object), it draws 450mA. The prototype performed basic hand movements, including
grasping an object, lifting it, and placing it back down. Figure
\ref{fig:fig5} shows the object grasping in action with the prosthetic hand.
This new research area has the potential to revolutionize
controlled prosthetic devices, as the deployment of a vision-controlled system provides a global solution for a much larger
number of users than the personalized training methods
employed using EMG and EEG-based prosthetics.
For EMG and EEG-based prosthetics, the training time
necessary for the user to operate the prosthetic effectively and
to train the model to detect the signal correctly can be extensive.
This includes both the user becoming comfortable with the
system, including any fitting adjustments, the training of
machine learning models, and so on. Naturally, the training
time of the device must be compared against the performance
of that device once trained. Our prosthetic hand design
maintains an access control strategy while maintaining the
benefits of a capable, fully actuated prosthetic. 

\section{Conclusion}
\label{sec:conclusion}
The number of people afflicted by the loss of an upper limb
is already over 3.6 million and is expected to only grow in the
coming years. This market, already worth over \$6 billion, is
characterized by expensive, hard to use, and inaccessible
products. Furthermore, losing this kind of mobility is
devastating to a patient’s independence; regaining this
independence can be empowering. Our project with the new
camera-based control system brings the sophistication of
traditional prosthetics into the mainstream market. This new
control system requires far less training time than those used by
conventional prosthetics, with the potential also to be far more
reliable. This is a proof of concept. The next step is to develop a
method to attach the hand so that it can be fitted for different
ages, sizes, and lengths of the lower arm. Our unique
combination of features brings hope to amputees to regain a
fully functional hand.

\bibliographystyle{unsrt}  
\bibliography{references}

\end{document}